\documentclass{emulateapj}
\usepackage[latin1]{inputenc}
\usepackage{amsmath}
\usepackage{amsfonts}
\usepackage{amssymb}
\usepackage{subfigure}
\usepackage{graphicx}
\usepackage[backref,breaklinks,colorlinks,citecolor=blue]{hyperref}        
\usepackage[all]{hypcap}

\usepackage{natbib}
\begin{document}
\title{A Monte Carlo Radiation Transfer Study of Photospheric Emission
  in Gamma Ray Bursts} 
\author{Tyler Parsotan \& Davide Lazzati}
\affiliation{Department of Physics, Oregon State University, 301 Weniger Hall, Corvallis, OR 97331, U.S.A.}

\begin{abstract}
  We present the analysis of photospheric emission for a set of hydrodynamic
  simulations of long duration gamma-ray burst jets from massive
  compact stars. The results are obtained by using the Monte Carlo
  Radiation Transfer code (MCRaT) to simulate thermal photons
  scattering through the collimated outflows. MCRaT
  allows us to study explicitly the time evolution of the photosphere
  within the photospheric region, as well as the gradual decoupling
  of the photon and matter counterparts of the jet. The results of the
  radiation transfer simulations are also used to construct light
  curves and time resolved spectra at various viewing angles, which
  are then used to make comparisons with observed data, and outline
  the agreement and strain points between the photospheric model and
  long duration gamma-ray burst observations. We find that our fitted
  time resolved spectral Band $\beta$ parameters are in agreement with
  observations, even though we do not consider the effects of
  non-thermal particles. Finally, the results are found to be
  consistent with the Yonetoku correlation, but bear some strain with
  the Amati correlation. 
\end{abstract}

\maketitle

\section{Introduction}
Long duration Gamma Ray Bursts (LGRBs) are some of the most energetic
events in the universe, emitting up to $10^{53}$ ergs of energy in the
first few tens of seconds \cite{Kulkarni_GRB_energy}. The progenitors for
these events are core collapse supernovae where the collapse of the
iron core into a supermassive object causes a relativistic jet to form
and flow through, and out of, the collapsing star. The prompt emission
that is produced by the jet can be explained by the synchrotron shock
model (SSM) \citep{SSM_REES_MES} and the photospheric emission model
\citep{REES_MES_dissipative_photosphere}. 

SSM considers the radiation that is produced by shells of varying
speeds colliding with one another outside of the photosphere. These
collisions cause magnetic fields and non-thermal particles to form
which then produce the radiation. This model is able to describe
various observations of LGRBs such as the non-thermal spectrum and the
light curve variability; however, the SSM is in tension with various
observational relations such as the Amati, Yonetoku and Golenetskii
Correlations \citep{Amati,Yonetoku,Golenetskii} as well as the
spectral indicies of the observed LGRB spectra \citep{SSM_index_prob}.

The photospheric model describes the radiation that is produced deep
in the jet, where the opacity is extremely high. The radiation
interacts heavily with the matter in the jet which affects the
spectrum more than the process that initially creates the radiation.
The trapped radiation is then released when the jet becomes
transparent to the radiation propagating through it. The photospheric
model is unable to naturally form a spectrum with non-thermal low and
high energy tails; however, the idea of the fuzzy photosphere
\citep{Peer_fuzzy_photosphere,Beloborodov_fuzzy_photosphere} and
sub-photospheric dissipation can cure the model's inability to form a
non-thermal high energy tail \citep{Atul}. 
{{\cite{vurm_radiation} also showed that inclusion of a variety of dissipation mechanisms can produce the observationally expected non-thermal low energy tail, although this has yet to be shown when considering a realistic GRB jet.}}
Opposing the SSM, a resounding success of the photospheric model is the model's
ability to reproduce all the observational correlations
\citep{diego_lazzati_variable_grb, lazzati_photopshere}.

Previous studies of the photospheric model have considered highly
detailed jet structures but assumed that the radiation and matter are
perfectly coupled until the photosphere, whereafter the two components
of the jet no longer interact with one another
\citep{lazzati_variable_photosphere}. Alternatively, other studies
have considered rigorous radiation transfer calculations on simplified
analytical outflows
\citep{ito_stratified_jets,ito_polarization,Lundman_photopsheric,Lundman_polarization,
  vurm_radiation}.

The bridge between complex jet structures and self consistent
radiation treatments has started to be built with
\citeauthor{Ito_3D_RHD}'s (\citeyear{Ito_3D_RHD}) investigation into
the radiation signature of a precessing jet with the use of a Monte
Carlo treatment of the radiation propagating through the jet. 
\cite{MCRaT} then proposed an independent algorithm, called MCRaT, which was tested on spherical and cylindrical 
outflows and then applied to a single 2D LGRB simulation. 

This paper continues along that path and presents the results of the MCRaT code
on an ensemble of special relativistic hydrodynamical simulations of LGRBs and what
the implications are for understanding these phenomena using the
photospheric model \citep{MCRaT}. This paper is structured such that
\autoref{methods} outlines the methods used in the analysis of the
MCRaT data, \autoref{results} discusses the results, and \autoref{end}
summarizes and discusses the implications of the results with theory
and observations.

\section{Methods}\label{methods}

As described in \cite{MCRaT}, MCRaT loads a frame of a FLASH
simulation and injects photons at an optical depth of $\sim 100$. The
code then propagates and Compton scatters each photon until the last
FLASH frame is reached. At this point, the code will restart by
injecting photons into the next frame from which the last set of
photons were injected. There are a total of $\sim 2.4\times 10^5$
photons in each simulation analyzed in this paper.

Once MCRaT has completed its run for a given hydrodynamical GRB
simulation, the results can be used to construct theoretical light
curves and time resolved spectra, much like what may be observed by
FERMI \citep{FERMI}. To this aim, we insert a virtual detector at a
suitable distance from the central engine and collect incoming photons
registering their arrival time and frequency. From the calculated
light curves and spectra, the MCRaT runs can then be compared to
individual bursts or to observational trends such as the Yonetoku and
Amati correlations \citep{Yonetoku, Amati}.  Any photon that
propagates past the location of the virtual detector at the end of the
radiation transfer simulation is collected and analyzed to produce
light curves and time resolved spectra.  From the 4-momenta, each
photon's direction of propagation and energy can be calculated. The
propagation direction allows for the identification of which photons
would be observed at a given angle relative to the GRB jet axis. Since
each photon travels along a specific direction, we must specify a
range of angles within which we collect the photons. Throughout this
paper we will specify the average of the acceptance angle range as the
viewing angle, $\theta_\text{v}$, and will consistently use a
$\pm 0.5^\circ $ acceptance range; for example, a viewing angle of
$1^\circ$ implies that all photons within $0.5^\circ$, inclusive, and
$1.5^\circ$, exclusive, have been collected.

The detection time, $t_\text{detect}$, for any photon that propagates
past the location of the virtual detector at the end of the radiation
transfer simulation depends on the detected jet launching time
$t_\text{j}$, the lab time at which the detection is evaluated
$t_\text{real}$, and the photon displacement time $t_\text{p}$, i.e.,
the amount of time the photon has traveled past the detector:
\begin{equation}
t_\text{detect}=t_\text{real}-t_\text{j}-t_p 
\end{equation}
To compute the detected jet launching time $t_\text{j}$, we consider a
virtual photon being emitted by the central engine at the time when
the jet is launched. This virtual photon is detected by the virtual
detector at $t_\text{j}=r_\text{d}/c$ which depends on the virtual
detector's distance from the central engine $r_\text{d}$.  In order
to know where to place the virtual detector, we must be able to
identify the location of the photosphere for each viewing angle. To do
this, we follow the method of \cite{MCRaT} to obtain a plot of matter
and photon temperature versus radial position from the jet.
In each FLASH frame, for a given set of photons, binned by detection
time, we calculate the average photon temperature and then identify
the FLASH grid points nearest to each photon and calculate the average
of the temperatures at each FLASH grid point. {  The temperature of the
photon in the fluid frame, $T_{\mathrm{ph}}$, is calculated as
\begin{equation}
T_{\mathrm{ph}}=\frac{h\nu}{3k}
\end{equation}
where h is Planck's constant, $\nu$ is the photon comoving frequency, $k$ is the
Boltzmann constant, and the factor of 3 comes from the photons being in Wien equilibrium with the matter \citep{spectral_peak_belo}. The fluid frame temperature of the matter in the FLASH simulation
grid, $T_{\mathrm{m}}$, is calculated as 
\begin{equation} \label{eq:3}
T_{\mathrm{m}}=\big( \frac{3p}{a} \big)^\frac{1}{4}  
\end{equation}
where $p$ is the pressure, and $a$ is the radiation density constant. This equation conforms with the assumption in the FLASH hydrodynamic simulations that the adiabatic index of the fluid is $4/3$; this assumption becomes an approximation either when the fluid temperature is non-relativistic or the radiation is not a blackbody. }

The plot of temperature versus radius, as discussed in
\autoref{results}, ideally, shows the average photon temperature { in the fluid frame}
approaching an asymptotic temperature. The distance at which the
average photon temperature remains constant is where the edge of the
photospheric region lies, and the virtual detector should be placed at
this distance or at a larger one. In practice, since this distance is
time dependent, we place our detector at the largest value for each
viewing geometry.

Once the detection time and angle are calculated for each photon, the
light curve can be constructed by collecting the photons that
propagate within a given angle range, corresponding to
$\theta_\text{v} \pm 0.5^\circ $, and then binned into 1 second bins,
based on the arrival time of the photon. To convert the light curve
from counts to luminosity the sum of the photons' weights times
energies are calculated. The weights are determined following equation
1 in \cite{MCRaT}.

The time resolved spectra are obtained by binning the observed
photons, within a time interval of 1 second, into energy bins of
logarithmically increasing widths. The number of photons collected in
an energy bin can be converted into intensity by summing up each
photon's weight times energy and then dividing by the bin width.

The uncertainty for each point in the spectrum is derived assuming a
Poisson distribution of counts. This makes the error associated with
an energy bin the intensity at that energy bin divided by the square
root of the number of photons collected in the energy bin. In order to
perform a $\chi^2$ minimization to do a spectral fit, we only fit the
energy bins that have at least 10 photons.  We fit a Band function
\citep{Band} to the time resolved spectra in order to
derive: the low and high energy spectral indices, $\alpha$ and $\beta$,
and the peak energy, $E_{pk}=E_o (2+\alpha)$, where $E_o$ is the break
energy. This is the definition used in \cite{Amati} and
\cite{FERMI}. On the other hand, spectra can also be fit by the simpler
comptonized (COMP) spectrum model (\cite{FERMI, BATSE_catalog}).
This is the model in which the Band function's
$\beta \rightarrow -\infty$. 

In order to determine which function is statistically significant in
fitting the data, we conduct an F-test. If the resulting p-value of
the F-test is less than 5\% -- the conventional false positive
acceptance rate -- the Band function, with an extra free parameter,
provides a statistically superior fit and is used. If the p-value is
greater than 5\%, then the simpler COMP fit is utilized to fit the
spectrum. Furthermore, spectra where the fitted values of $\alpha$,
$\beta$, and $E_{pk}$ were not well constrained were excluded from the
analysis.

\section{Results} \label{results}

\begin{table}
\begin{center}
\caption{Simulation Set Parameters}
\label{sims}
\begin{tabular}{lccc}
\hline Simulation Name & Progenitor & Jet Luminosity (erg/s) & $\Gamma_\infty$ \tablenotemark{a} \\ 
\hline 16OI & 16OI & $5.33 \times 10^{50}$ & 400  \\ 
 35OB & 35OB& $5.33 \times 10^{50}$ & 400  \\ 
 16TI & 16TI & $5.33 \times 10^{50}$ & 400  \\ 
 16TI.e150 & 16TI & $1 \times 10^{50}$ & 400  \\ 
 16TI.e150.g100 &  16TI & $1 \times 10^{50}$ & 100 \\ 
\hline 
\end{tabular}
\tablenotetext{1}{Asymptotic Lorentz factor}
\end{center}
\end{table}

In this section we present the results obtained by running MCRaT on
various FLASH simulations of relativistic jets from several LGRB
progenitor from \cite{Woosley_Heger}. These are the same simulations
used by \cite{lazzati_photopshere}. They have a jet injection radius
of $1\times 10^9$ cm, an initial lorentz factor of 5, an opening angle
of $10^\circ$, and the engine was turned on for 100 s. The differences
among the simulations are outlined in \autoref{sims}.  The MCRaT
simulations were run for at least 50 simulation seconds.

\begin{figure*}[t!]
\subfigure{\includegraphics[width=0.33\textwidth]{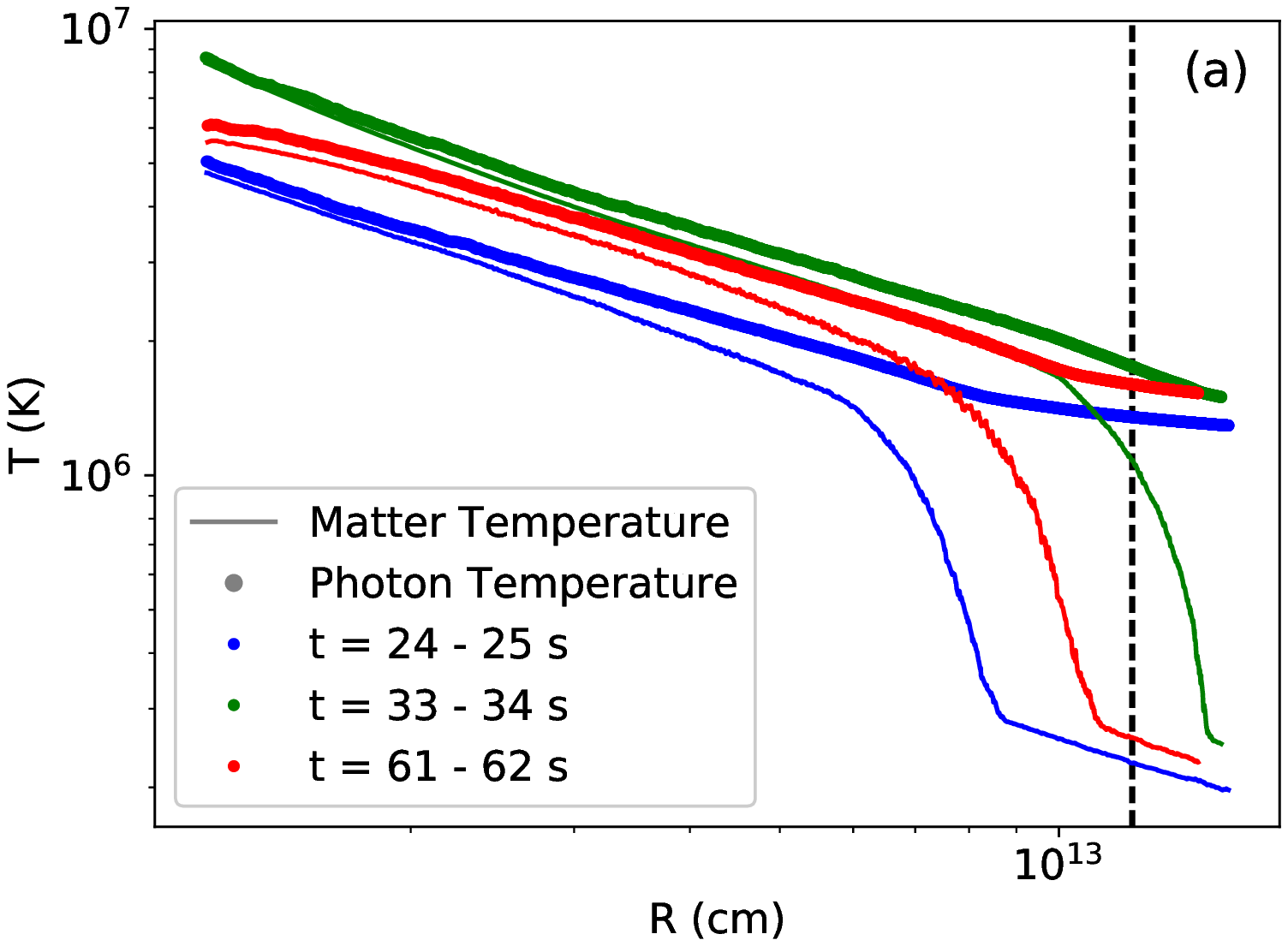}} 
\subfigure{\includegraphics[width=0.33\textwidth]{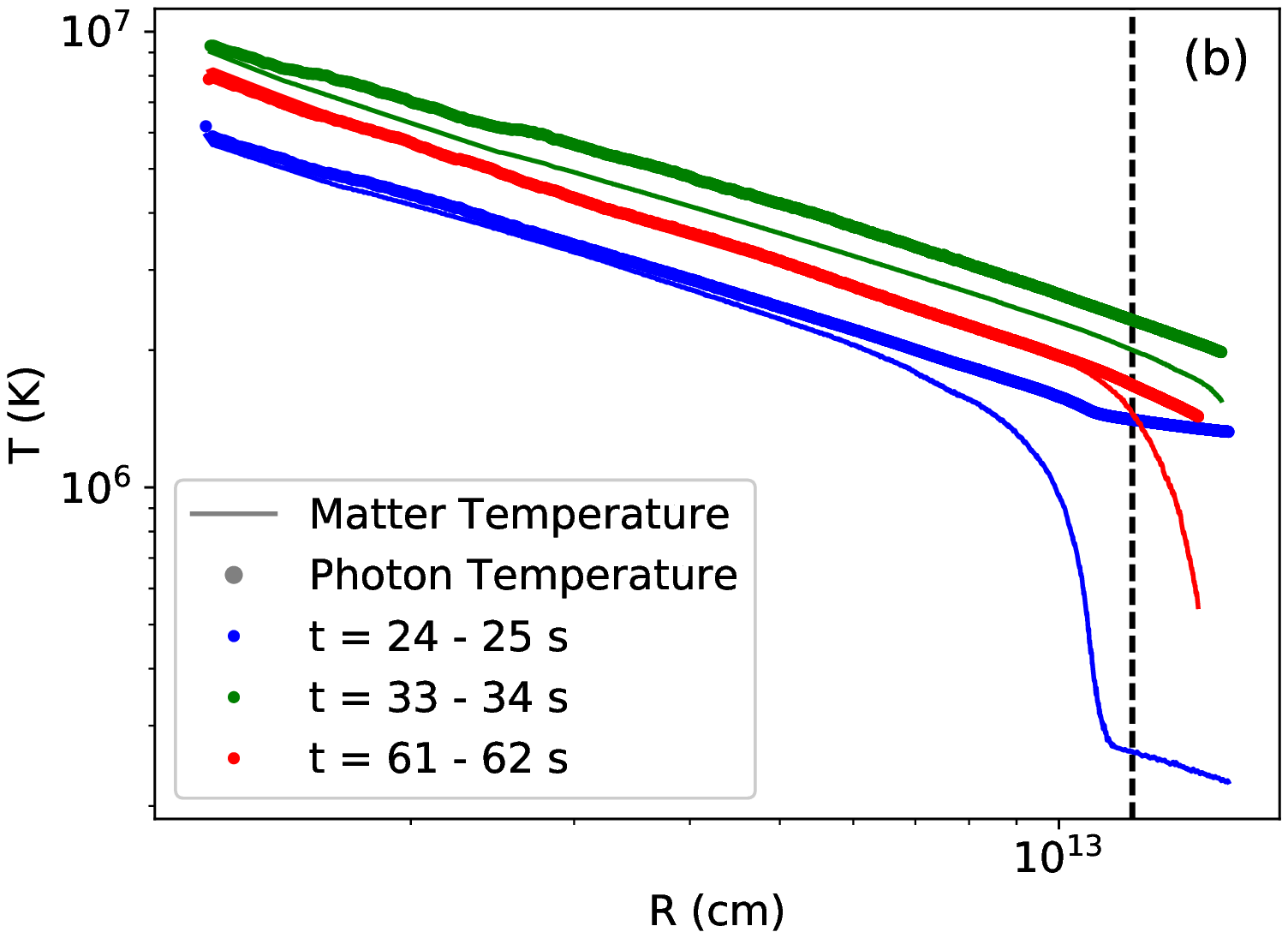}} 
\subfigure{\includegraphics[width=0.33\textwidth]{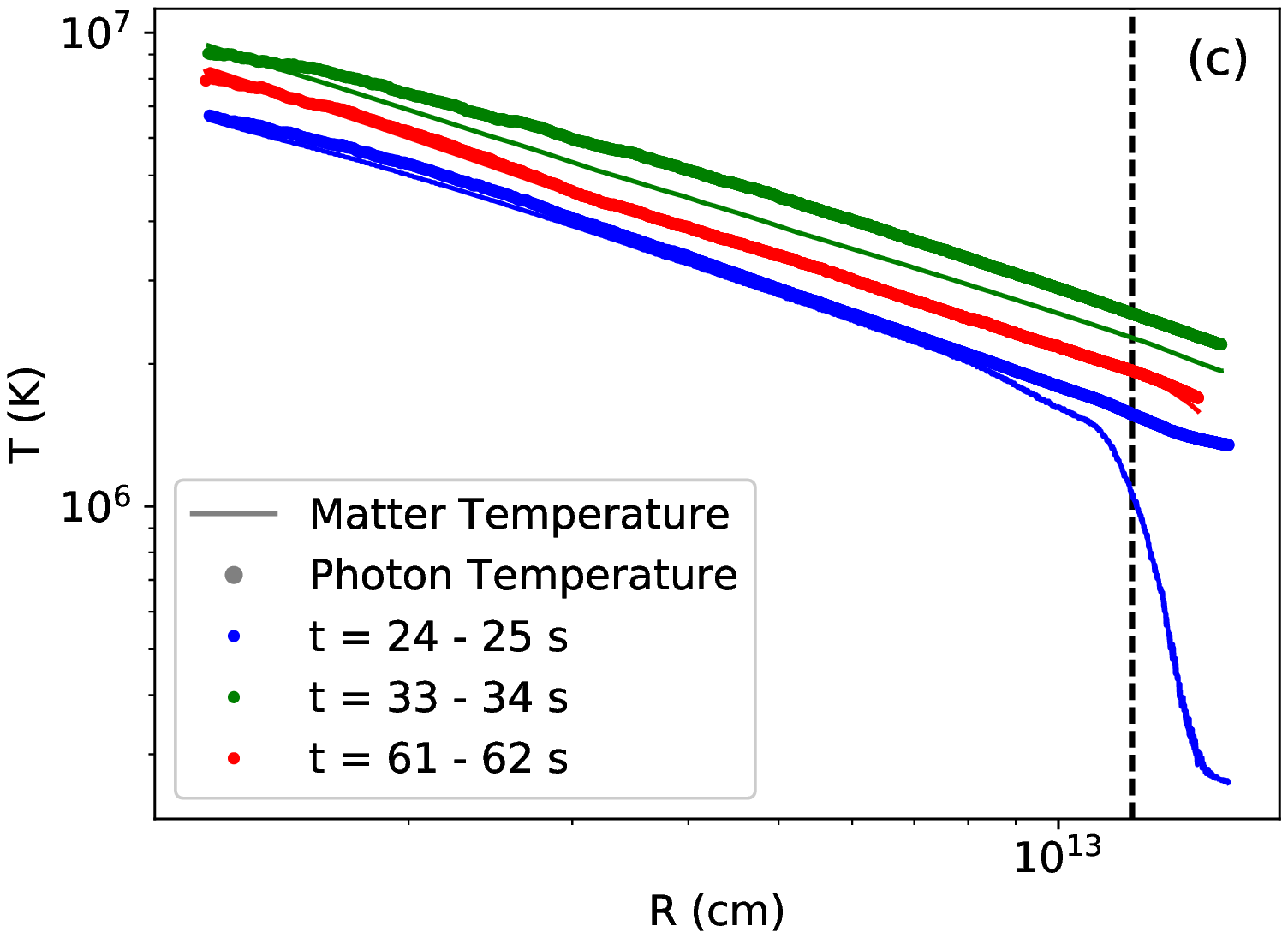}}
\caption{ Radiation and matter
  temperatures { in the fluid frame} as a function of distance from the central engine for
  the 35OB progenitor at viewing angles of $1^\circ$, $2^\circ$ and $3^\circ$, for figures (a), (b), and (c) respectively. The various
  colors represent photons which have been detected at different
  times. The circle markers represent the average photon temperature
  and the line is the average temperature of the FLASH grid points
  nearest to the photons of interest. The vertical dashed black line
  represents the location of the virtual detector at
  $1.2 \times 10^{13}$ cm.}
\label{35OB_temps}
\end{figure*}

\begin{figure}[]
\plotone{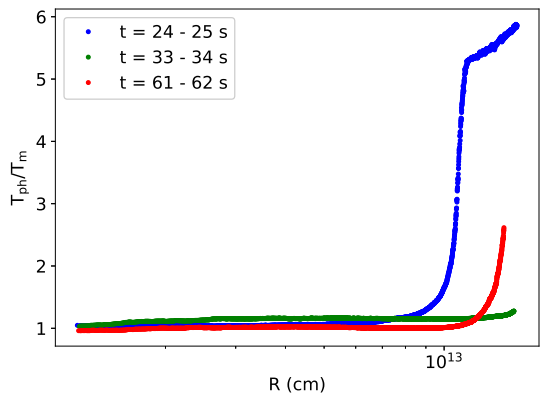} 
\caption{The ratio of the { comoving} photon temperature, $T_{\mathrm{ph}}$, to
  the { comoving} matter temperature, $T_{\mathrm{m}}$, is plotted for photons
  detected at differing times since the jet launch as shown in
  Figure 1(b). The ratios begin to increase as the photons decouple
  from the jet. As such, each of the times show a different
  divergence point.}
\label{temp_ratios}
\end{figure}

\subsection{Decoupling of the Photons and the Matter}

As described above, we use the comparison between the plasma and
radiation temperatures to determine where the virtual detector needs to
be placed. The plasma temperatures are read directly from the FLASH
simulations, while the radiation temperature is computed by MCRaT.

{ \autoref{35OB_temps} shows the average fluid frame photon and matter temperatures
as a function of distance from the central engine for the 35OB
simulation at various viewing angles. For all the curves shown, the
photons and matter start out coupled. However, the coupling
drastically changes as the temperatures of the photon and matter begin
to diverge from one another. While the matter continues in its cooling
behavior, the photons cooling slows as the Compton coupling between
matter and radiation weakens. Only when the decoupling becomes
complete does the temperature of the photons becomes constant. Notice how
the photons stay coupled to the matter for a larger distance at high
viewing angles. This can be understood as due to the baryon
entrainment in the jet along the boundary. Baryon entrainment
increases opacity by both increasing the density of electrons and by
reducing the Lorentz factor.}

Each sub figure in \autoref{35OB_temps} shows that the photosphere is
not a static surface in space, but rather an evolving surface that can
vary quite dramatically. Between the three times shown for each of the
viewing angles, the photosphere moves from being within the domain of
our simulation at $t=3$s, to being out of the domain of our simulation
at $t=30$s, and back to being relatively far back in the jet at
$t=50$s. Hence, we refer to this region in which the photosphere moves
as the ``photospheric region''. In the small temporal and spatial
domain of the simulations, we do not definitively reach the edge of
the photospheric region at high viewing angles,
$\theta_\text{v} \geq 3^\circ$. This effect is caused by the higher
densities at $\theta_\text{v} \geq 3^\circ$, compared to the densities
at angles closer to the jet axis. As a result of not reaching the
photosphere for some populations of photons, our energies should be
taken as an upper limit for the cases of high viewing angle. In order
to get to the photosphere at high viewing angles, we would need to run
FLASH simulations of larger domain. These, however, can be carried out
only at the price of reducing the resolution (e.g., \cite{Ito_3D_RHD}). 

The changing location of the photosphere for a given GRB simulation is
also shown in \autoref{temp_ratios}, where the ratios are plotted for
the three time periods shown in Figure 1b at a viewing angle of
$2^\circ$. Each curve begins to turn away from a ratio of 1 at a
different distance, indicative of the fact that the jet and the matter
decouple at different distances for different times within the
simulated jet.  These characteristics of the temperature plots are
consistent for all the simulations in our set.

\subsection{Light Curves and $E_{pk}$}

\begin{figure*}[t!]
\subfigure{\includegraphics[width=0.33\textwidth]{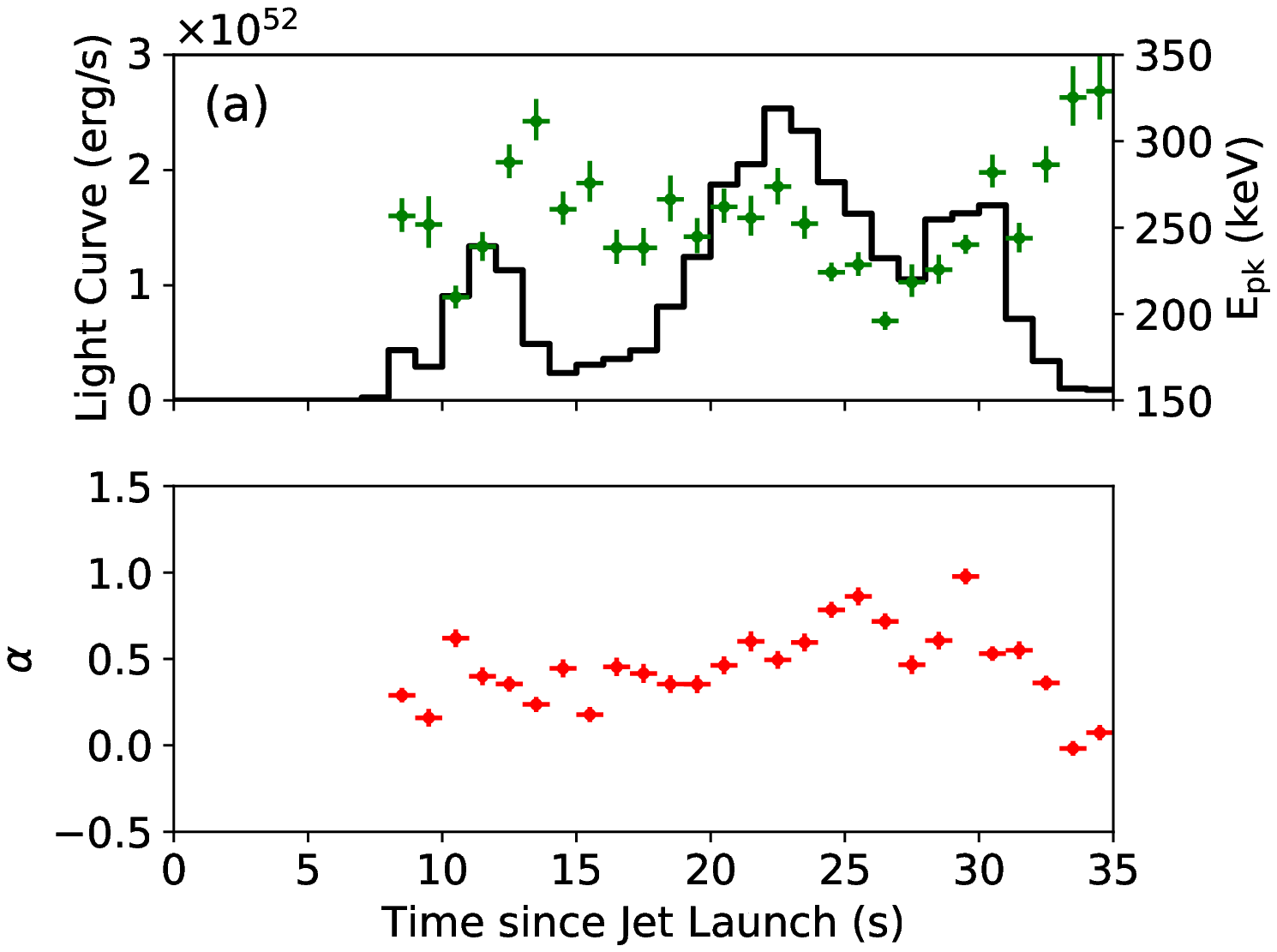}} 
\subfigure{\includegraphics[width=0.33\textwidth]{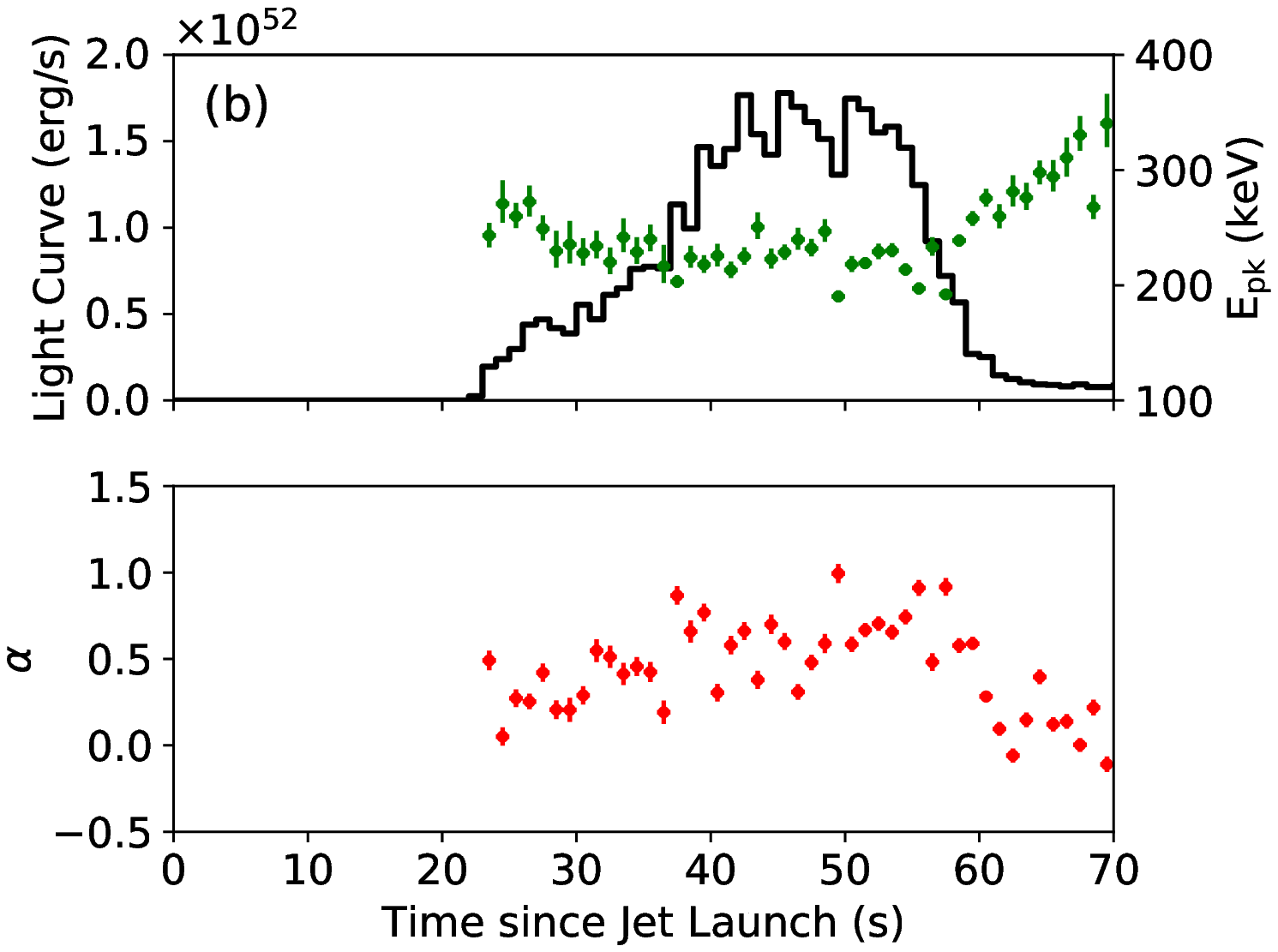}} 
\subfigure{\includegraphics[width=0.33\textwidth]{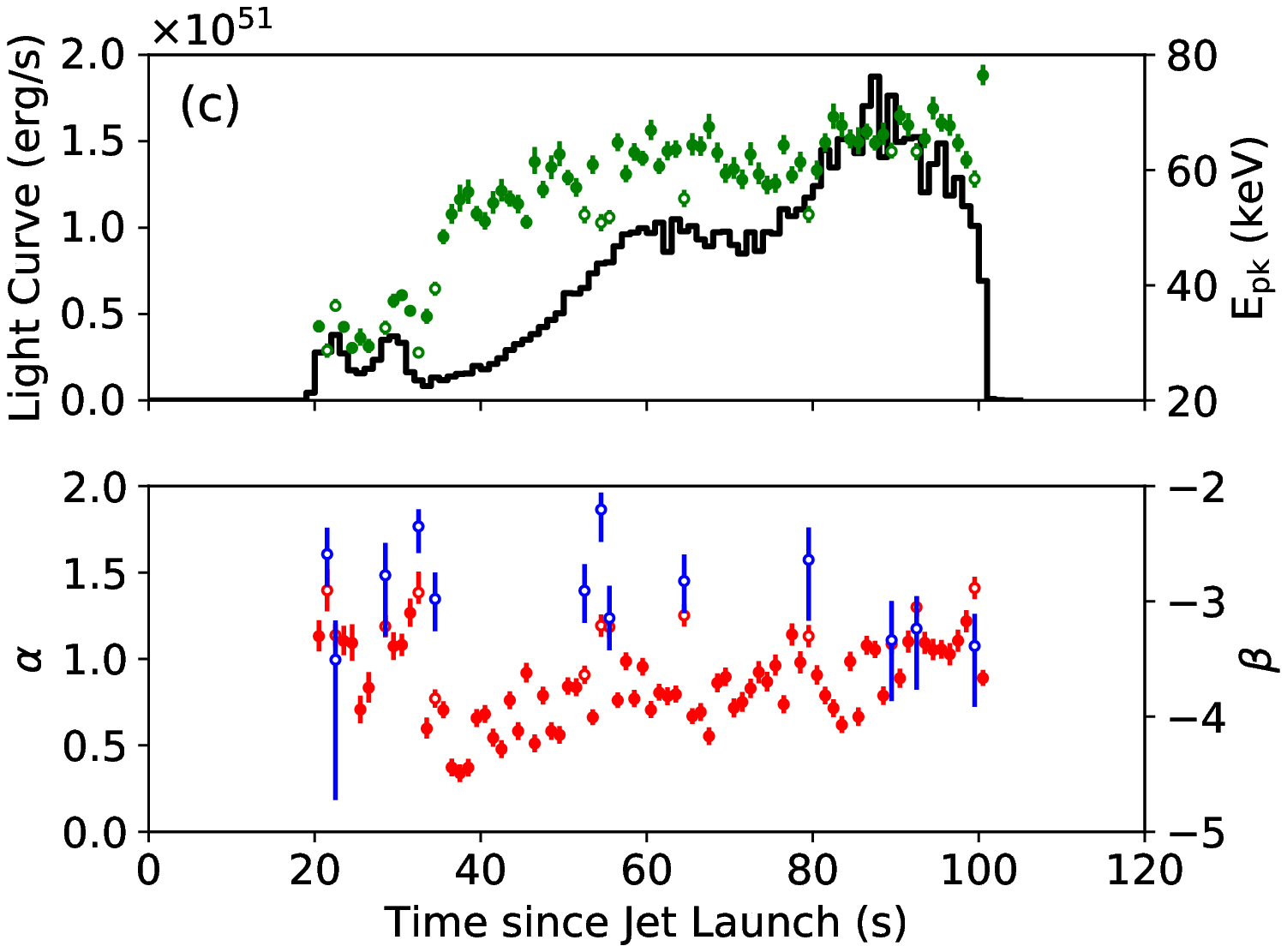}}
\caption{ Light curves from various simulations
  and viewing angles. Figure (a) shows the light curve of the 16OI progenitor at $\theta_\text{v} =1^\circ$, figure (b) shows the 35OB progenitor at $\theta_\text{v} =1^\circ$, and figure (c) shows the 16TI.e150.g100 light curve at $\theta_\text{v} =3^\circ$. These plots show how the various simulations can
  reproduce: a hard-to-soft tracking of $E_{pk}$, in (a),
  an anti-correlation between the luminosity and $E_{pk}$, in
  (b), and tracking between the $E_{pk}$ and the
  luminosity, as shown in (c). }
\label{various_light_curves}
\end{figure*}

\autoref{various_light_curves} show synthetic, time-resolved light
curves and spectra of the 16OI, 35OB, and 16TI.e150.g100 simulations
for various viewing angles (see \autoref{sims} for the details of the
simulations inputs). The top plots show the light curve as a black
line as well as the fitted peak energy as green markers with error
bars, as measured in a 1 second time bin. The bottom plots show the
temporal evolution of the fitted function parameters, where the low
energy parameter, $\alpha$, is in red and the high energy Band
parameter, $\beta$, is in blue. Many of the time resolved spectra were
satisfactorily fitted with COMP spectra, and are shown in solid
markers; other spectra were fitted with the Band function, if shown to
be statistically significant with an F-test, and are shown with
open markers.

The light curves are different from one simulation to the next, as
well as from one viewing angle to another within the same FLASH
simulation. The light curves exhibit diversity in their structure as
well as in the tracking between the fitted peak energy and the light
curve (or lack thereof). There are light curves for viewers at
relatively high viewing angles where a global hard-to-soft trend of
$E_{pk}$ is observed throughout the entire burst, while there are
other cases in which the peak energy tracks the light curve. Other
light curves do not display any clear hard-to-soft or tracking
behavior. These results are qualitatively analogous to the
observations made by FERMI (see Figure 6 in \cite{FERMI}).

In most cases, we see that viewing angles close to the jet axis,
$\theta_\text{v} = 1^\circ - 2^\circ$, exhibit either a global
hard-to-soft trend or a lack of an obvious trend. At higher viewing
angles $\theta_\text{v} = 3^\circ $, we typically see a tracking
behavior.

\subsection{Analysis of the Spectral Fits}

\begin{figure*}[t!]
\subfigure{\includegraphics[width=0.33\textwidth]{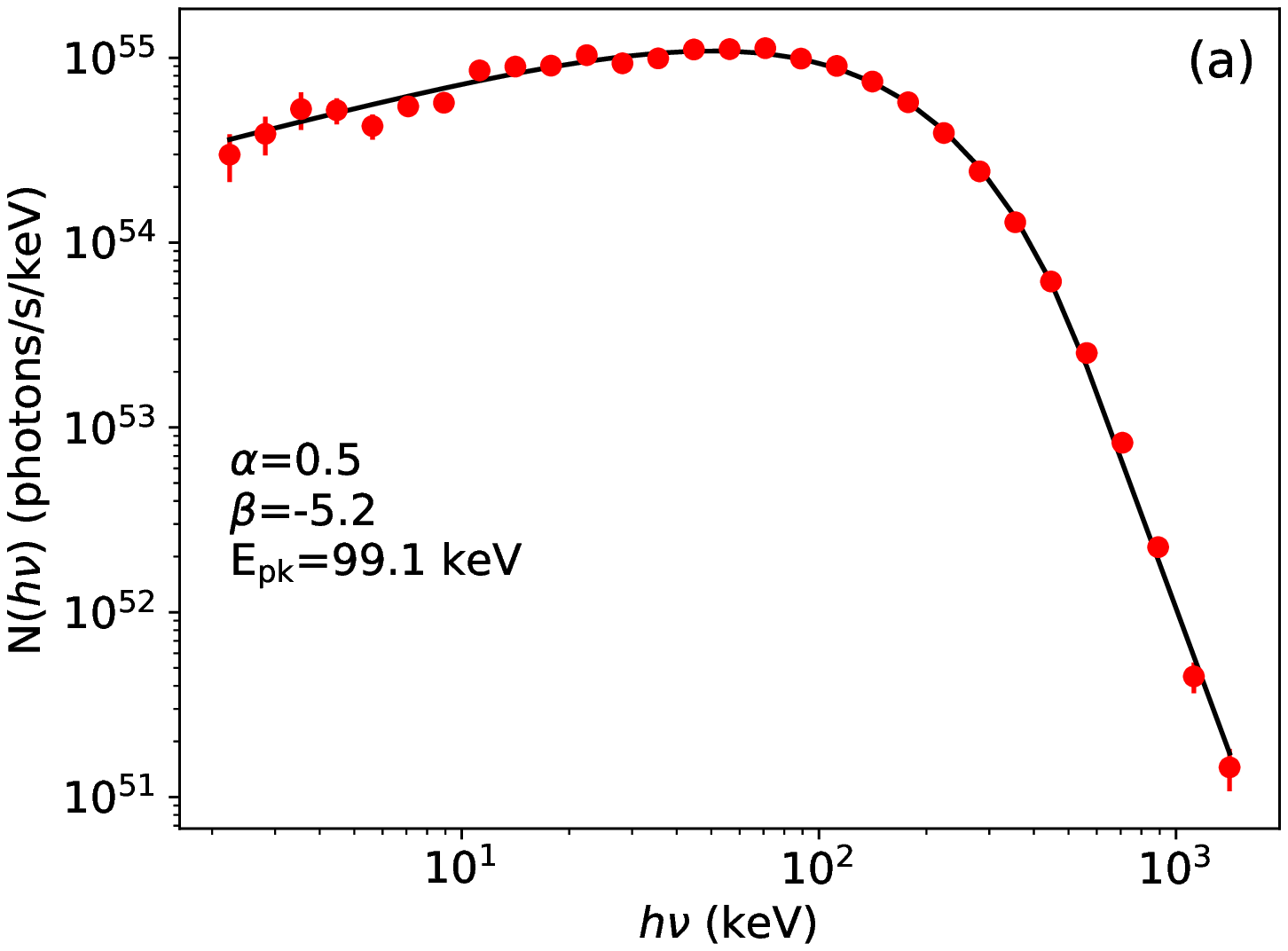}} 
\subfigure{\includegraphics[width=0.33\textwidth]{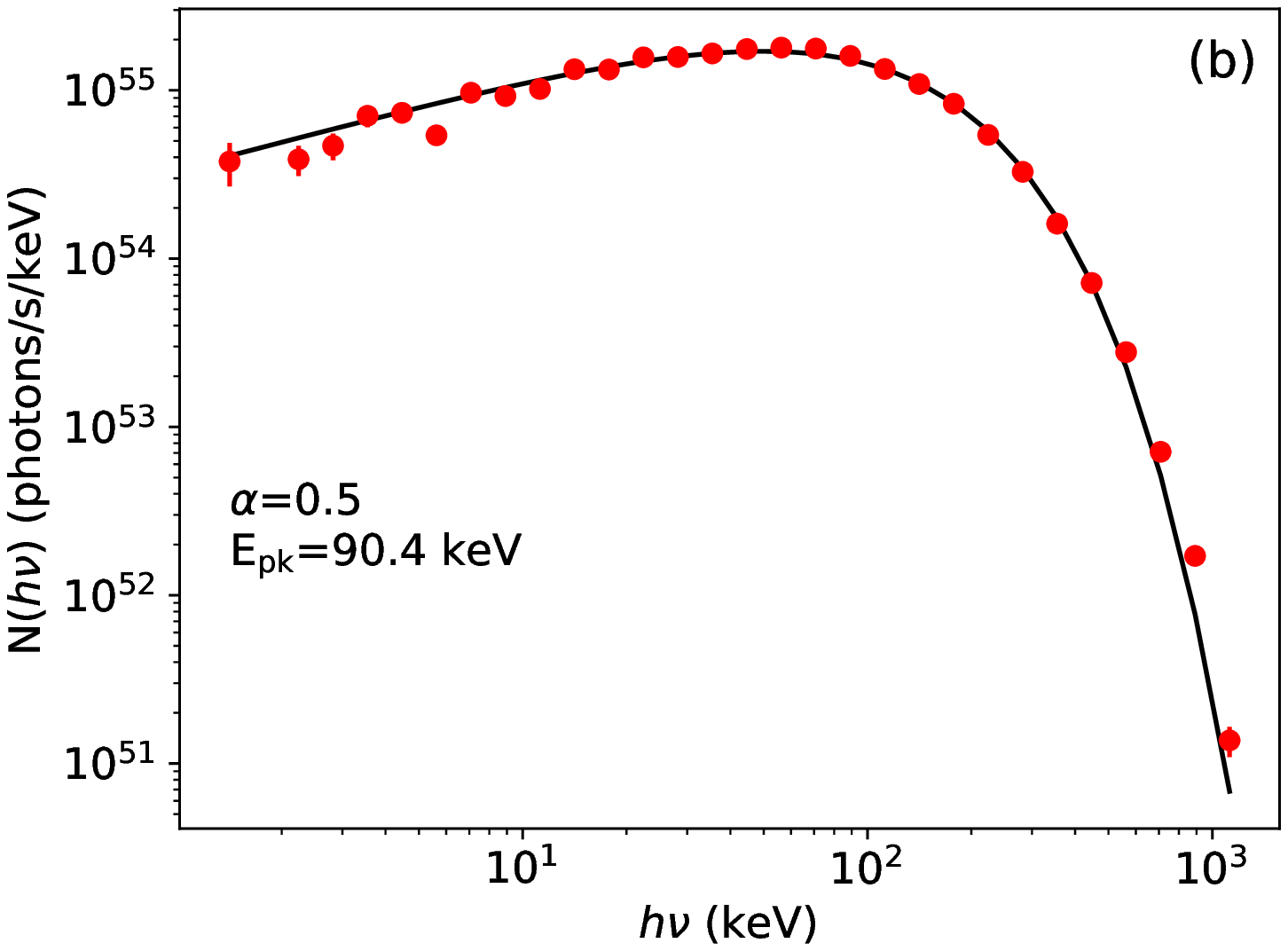}} 
\subfigure{\includegraphics[width=0.33\textwidth]{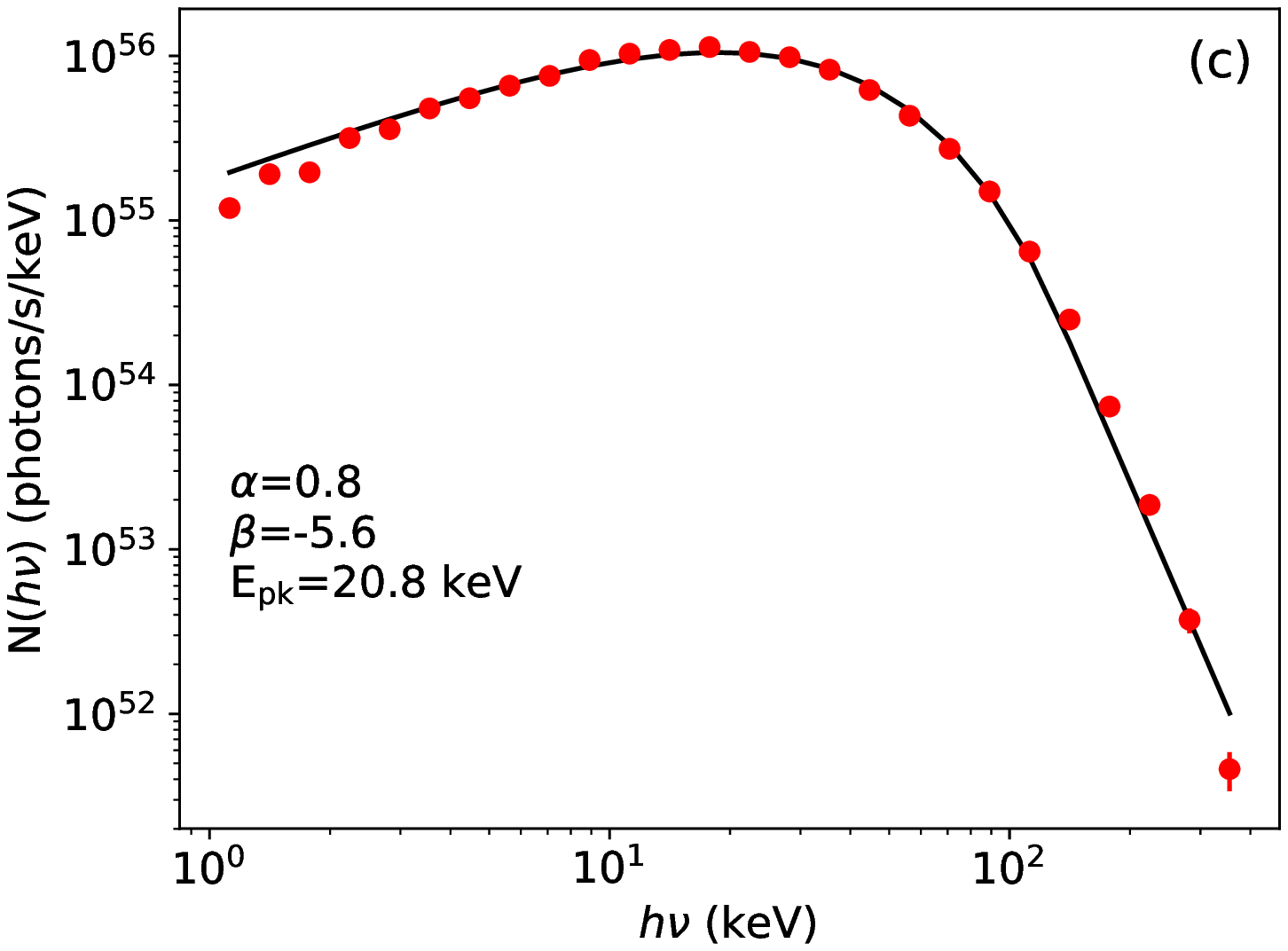}}
\caption{Plots (a), (b) and (c) show the time-integrated spectra of the 16OI, 35OB, and 16TI.e150.g100
  simulations respectively. The viewing angles for each simulation are the same as in \autoref{various_light_curves}. The best fit parameters for each spectra are
  included on each plot. The 16OI and 16TI.e150.g100 spectra are best
  fit with a Band spectrum and the 35OB spectrum is best fit with a
  COMP spectrum. }
\label{various_time_int_spectra}
\end{figure*}

\begin{figure*}[]
\subfigure{\label{all_params_a} \includegraphics[width=0.33\textwidth]{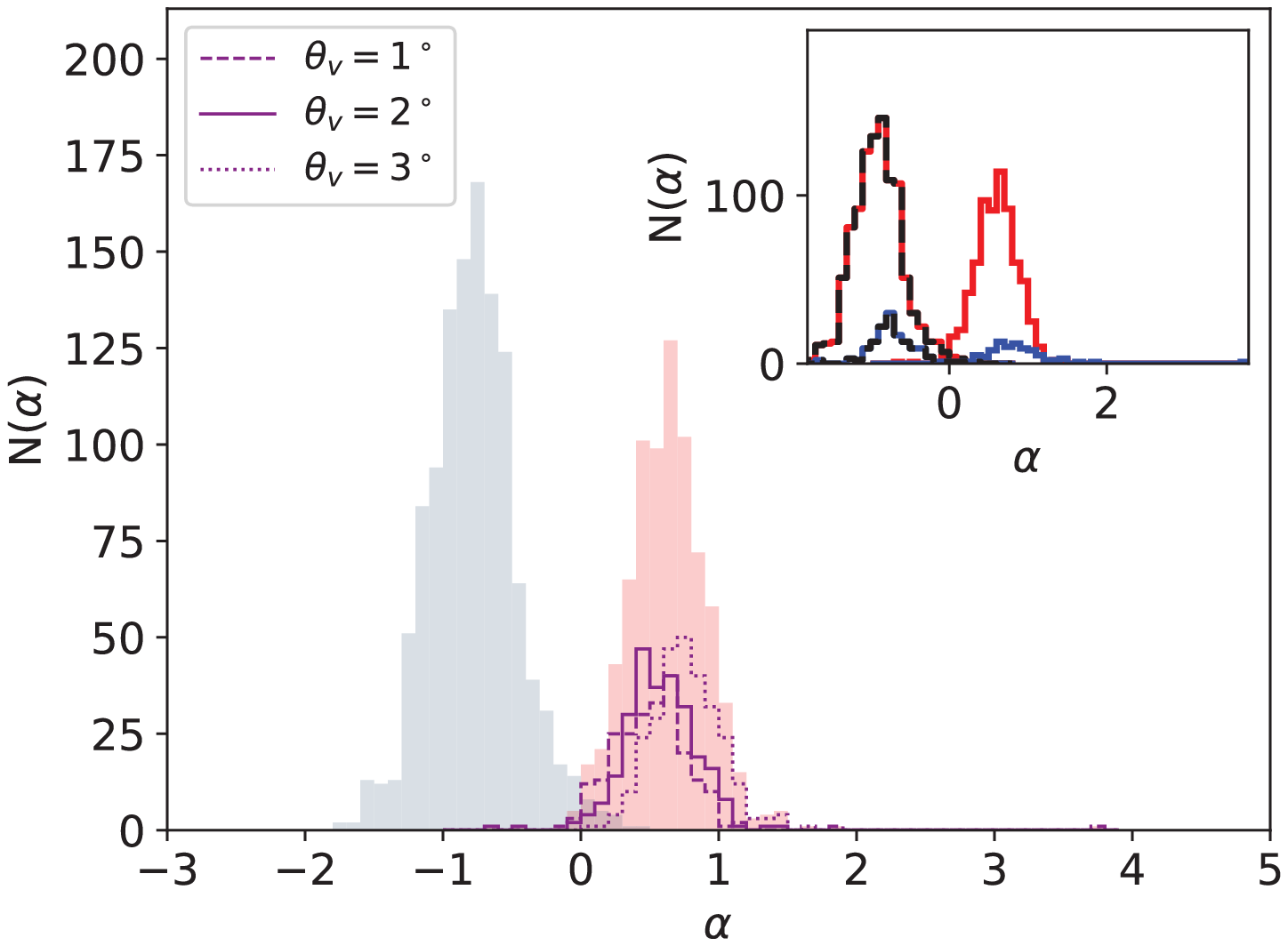}} 
\subfigure{\label{all_params_b} \includegraphics[width=0.33\textwidth]{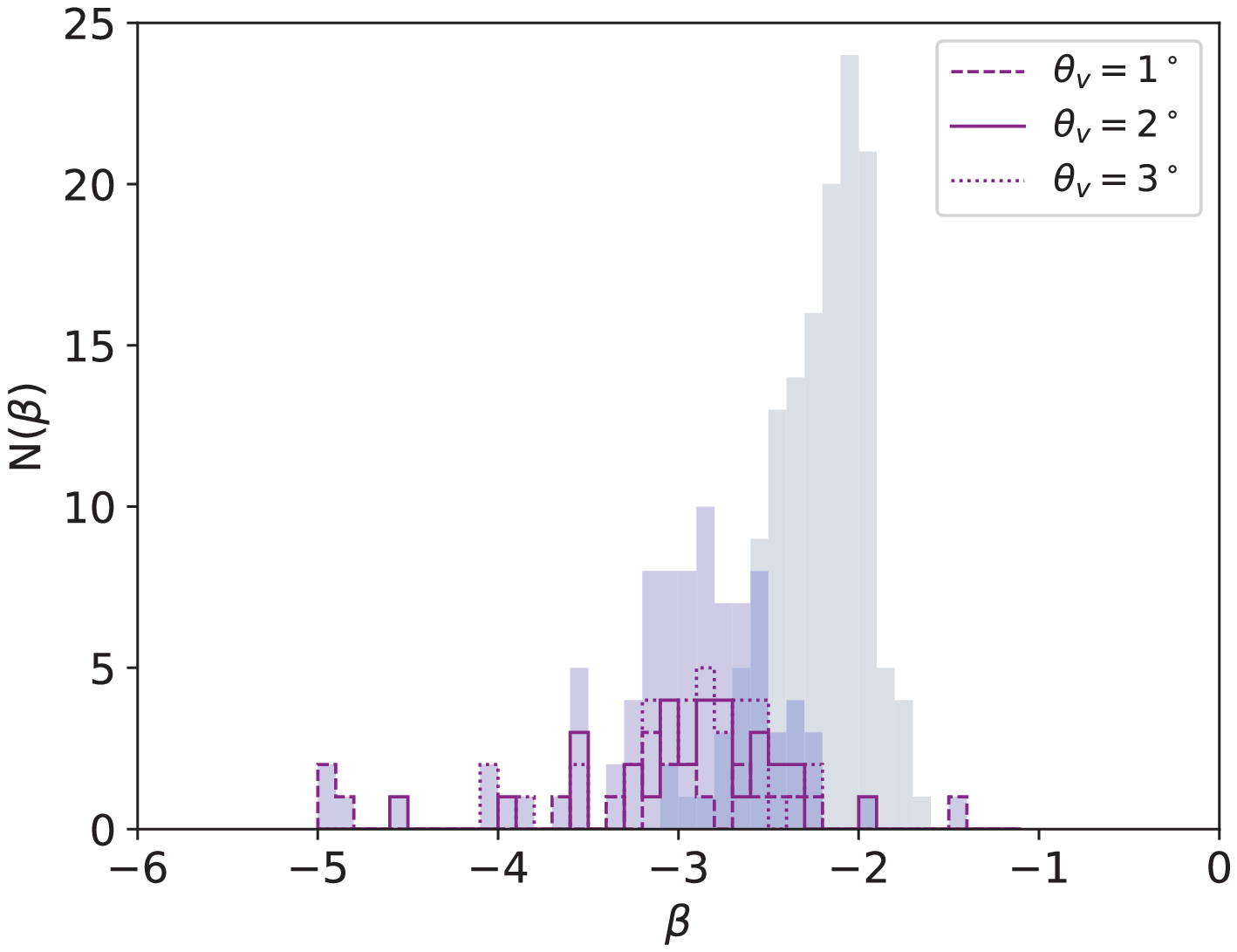}} 
\subfigure{\label{all_params_c} \includegraphics[width=0.33\textwidth]{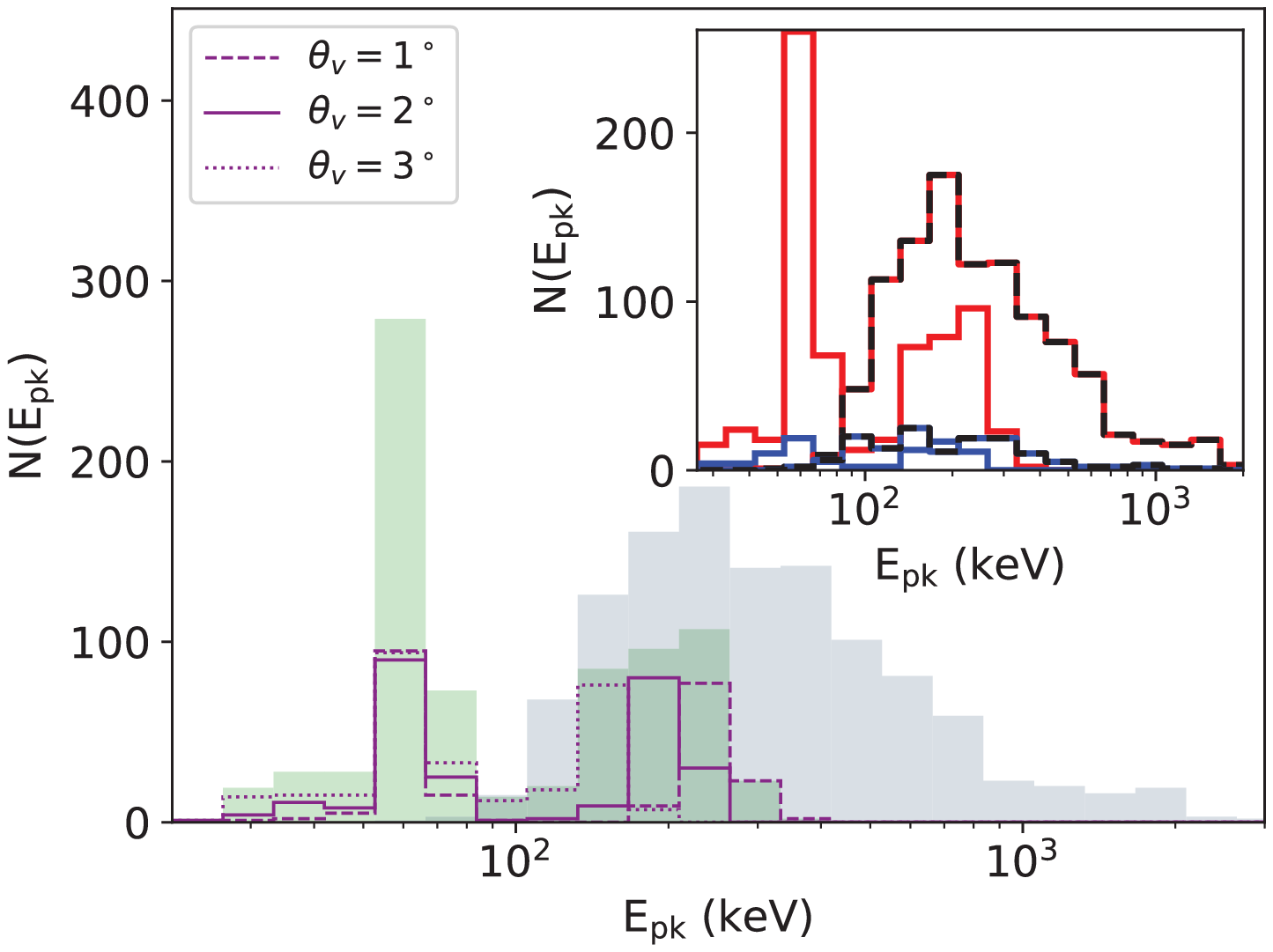}}
\caption{Distributions of the fitted Band and COMP Parameters for the
  total simulation set. The grey histograms are the fitted parameters
  from \cite{FERMI}. The distributions are separated by viewing angle
  and are shown by the purple dashed, solid, and dotted lines for
  $\theta_\text{v} = 1^\circ, 2^\circ$ and $3^\circ$,
  respectively. The inset plots show the distributions of the
  simulation set and the FERMI data set differentiated by whether the
  COMP or Band function was used to acquire the parameter, with the
  exception of the $\beta$ distribution since that is only defined for
  the Band function. The COMP is shown in red, the Band parameters are
  shown in blue, and the FERMI data is shown as a red and black dotted
  line or a blue and black dotted line, corresponding to a COMP or
  Band fit, respectively.}
\label{all_params}
\end{figure*}

The best fit parameters are collectively shown in
\autoref{all_params}, for the 16OI, 16TI, 35OB, 16TI.e150.g100, and
16TI.e150 progenitors. The parameters for the full simulation set are
grouped together and compared to the FERMI data, shown in grey. The
different distributions for each viewing angle are shown by the purple
dashed, solid, and dotted lines for
$\theta_\text{v} = 1^\circ, 2^\circ$ and $3^\circ$, respectively. The
inset plots, for $\alpha$ and $E_{pk}$, show how the distribution of
fitted parameters change between the COMP fits, in solid red, and the
Band fits, in solid blue. The FERMI data is also split into COMP and
Band fits which are represented by red and blue dotted lines,
respectively.

Looking at the distribution of fitted $\alpha$ parameters,
irregardless of the type of function that was used in the fit, it is
easy to see that our low energy spectral indices average around
$\sim 0.5$ while the FERMI distribution is clustered around
$\sim -0.75$. On the other hand, our distribution of $E_{pk}$ coincide
relatively well with FERMI's distribution of the peak energy for all
simulations except for the 16TI.e150.g100 and 16TI.e150
simulations. These simulations make up the large number of spectra
with energies less than $\sim 100$ keV. This is to be expected since
those simulations have lower engine luminosity.  Overall, the majority
of our spectra are well fit by the COMP spectrum which is to be
expected since we are solely considering Compton scattering. However,
we do have spectra where the Band function provides a statistically
superior fit. Analyzing our Band $\beta$ parameters and comparing them
to the FERMI distribution, we show that the observed Band $\beta$
parameter can be reproduced in a significant number of cases. This is
due solely to photons being upscattered near the photosphere, which is
a consequence of the fuzzy photosphere or ``photospheric region''
\citep{Peer_fuzzy_photosphere,Beloborodov_fuzzy_photosphere}.

\begin{figure}[]
\plotone{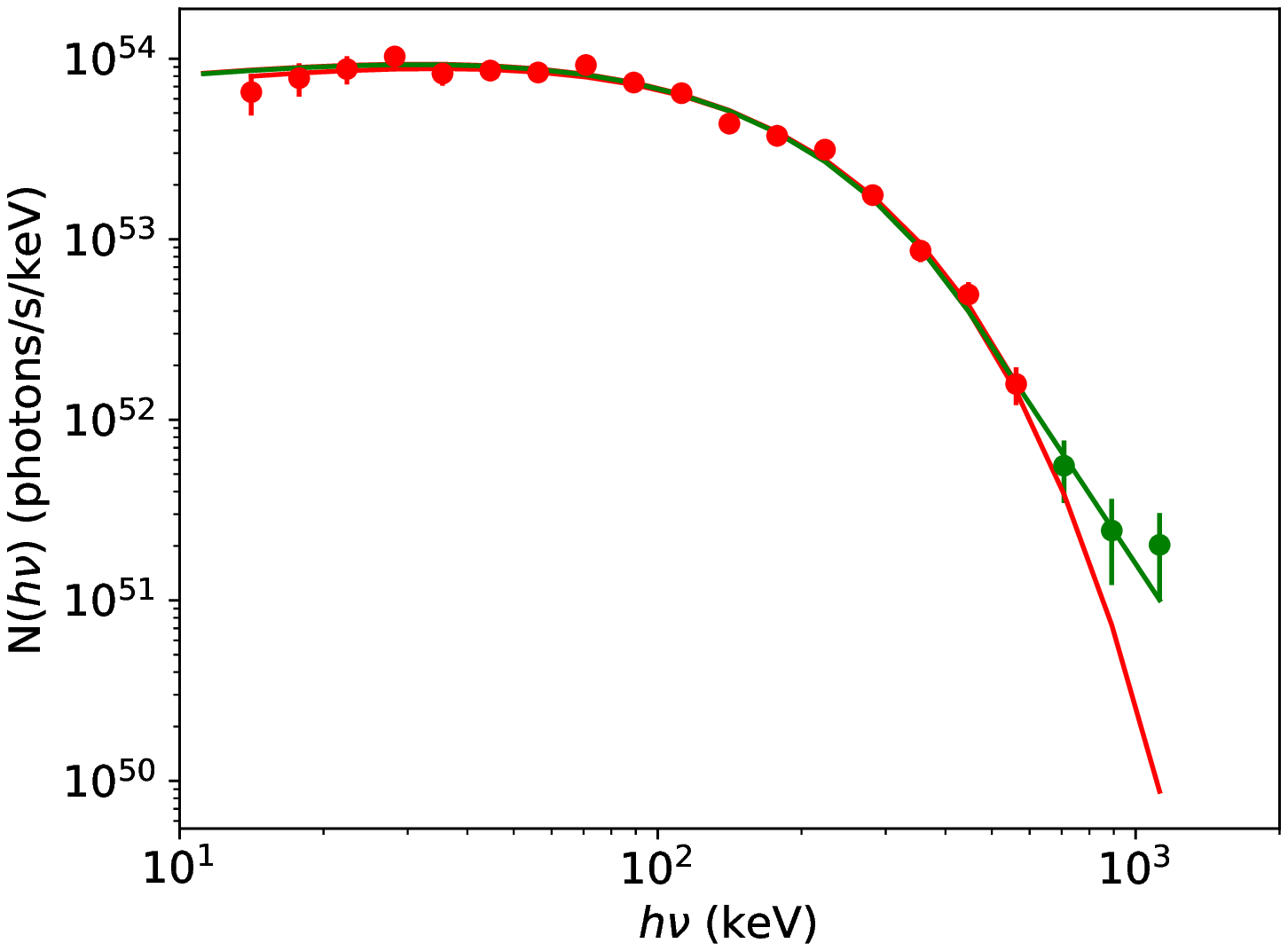} 
\caption{The points with $>$10 photons in their respective bins are
  shown in red with the COMP best fit plotted in red. The green points
  are bins that were excluded from the analysis due to low photon
  counts, however, if these bins are included in the fitting then the
  best fit function becomes the Band function, plotted as the green
  line.}
\label{comp_to_band}
\end{figure}

The preference for a COMP fit over a Band spectrum is not a unique
feature of our synthetic spectra. As a matter of fact, observed
time-resolved spectra also display a preference for a COMP fit with
69.1\% of the spectra being fit with the COMP model
(\citeauthor{FERMI} \citeyear{FERMI}). As \citeauthor{FERMI} explain,
the preference to the COMP model is due to low photon counts at high
energies, reducing the ability for a model to fit those high energy
bins well. Since our simulations emulate a typical GRB observation by
binning photons in time, space, and energy, our simulations suffer
from these low statistics as well. \autoref{comp_to_band} shows
spectral points that were not included in the original fit of the 16TI
time resolved spectrum, at $t=20-21$s, due to low photon counts; if
such rejected points are included in the curve fitting, the preferred
fit becomes the Band function instead of the COMP function. In order
to compensate for low photon counts we will need to inject more
photons in our simulations in the future; of course, this will be at
the expense of computational time.

It is enticing that only considering Compton scattering our synthetic
spectra can reproduce nearly all the observational features of GRB
prompt spectra, with the notable exception of the low energy index
$\alpha$. We show the time-integrated spectra for various simulations
and viewing angles in \autoref{various_time_int_spectra}. The
parameters for the best fit spectrum are shown on each
plot. Figure 4(a) is best fit with a Band function
although each time resolved spectrum is best fit with the COMP
spectrum, showing that various thermalized spectra can add together to
form a non-thermal spectrum. On the other hand, the time-integrated
spectrum of the 35OB simulation at $\theta_\text{v} = 1^\circ $, in
which every time resolved spectrum is best fit with the COMP spectrum,
is also best fit by the COMP function, as shown in
figure 4(b). The time-integrated spectrum of the
16TI.e150.g100 simulation at $\theta_\text{v} = 3^\circ $, which is
best fit with the Band function, as shown in
figure 4(c), is formed from time resolved
spectra which are best fit by both the Band and COMP functions.

Another important aspect of our results is that the calculation of
$E_{pk}$ is dependent on the fits for $\alpha$, which is greater than
what is observed by $\sim 1$. This causes our peak energies to be
somewhat artificially high. The issue of correctly reproducing the
$\alpha$ parameter is a well known issue in the photospheric model. It
can be fixed by including photon emitting processes, such as a
synchrotron component. As the spectrum becomes corrected, with a larger number of low energy photons producing the 
proper values of $\alpha$, we would expect $E_{pk}$ to decrease \citep{spectral_peak_belo}.

\subsection{Comparison to Observational Relations}

\begin{figure*}[]
\epsscale{1.10}
\plottwo{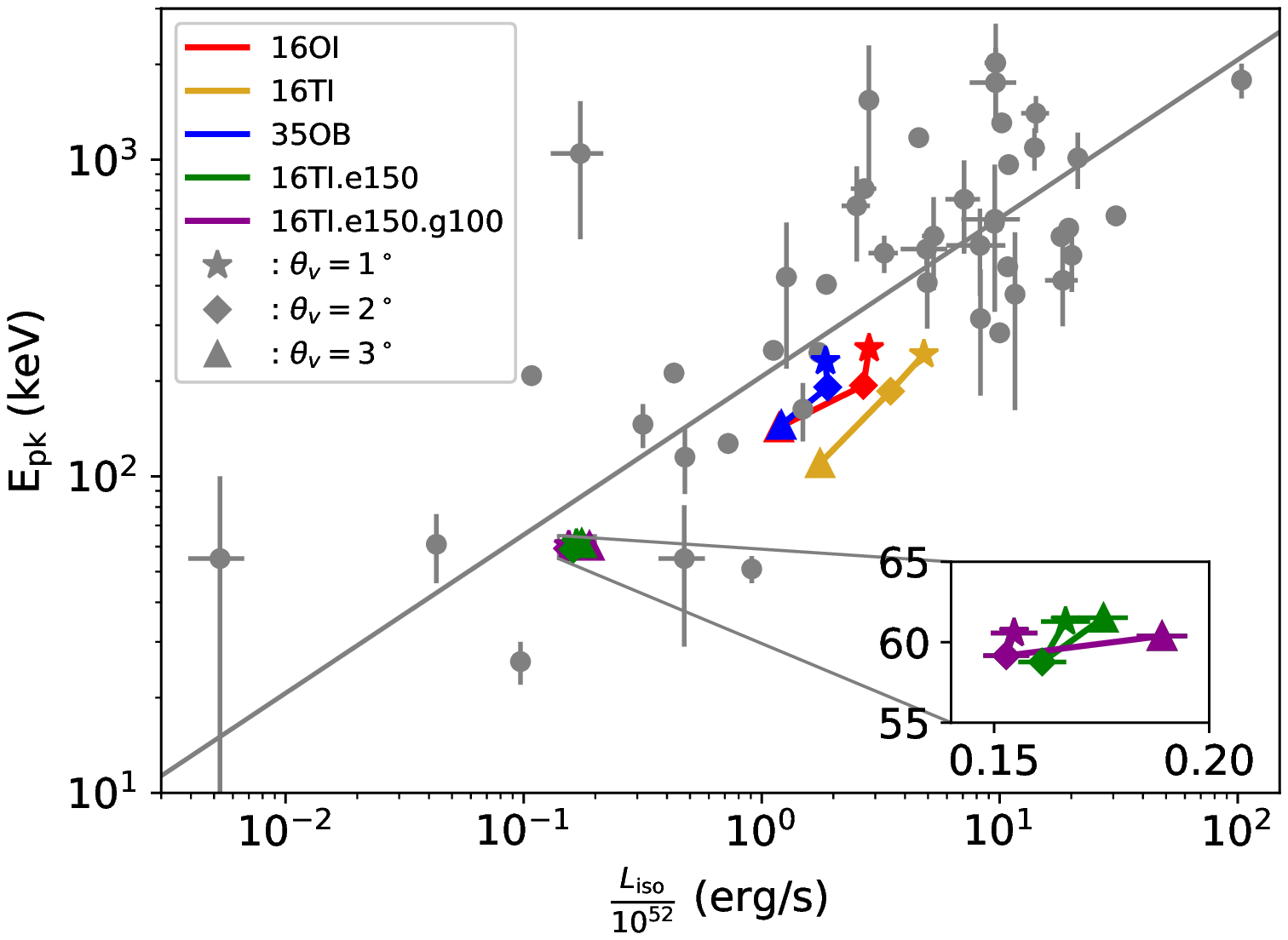}{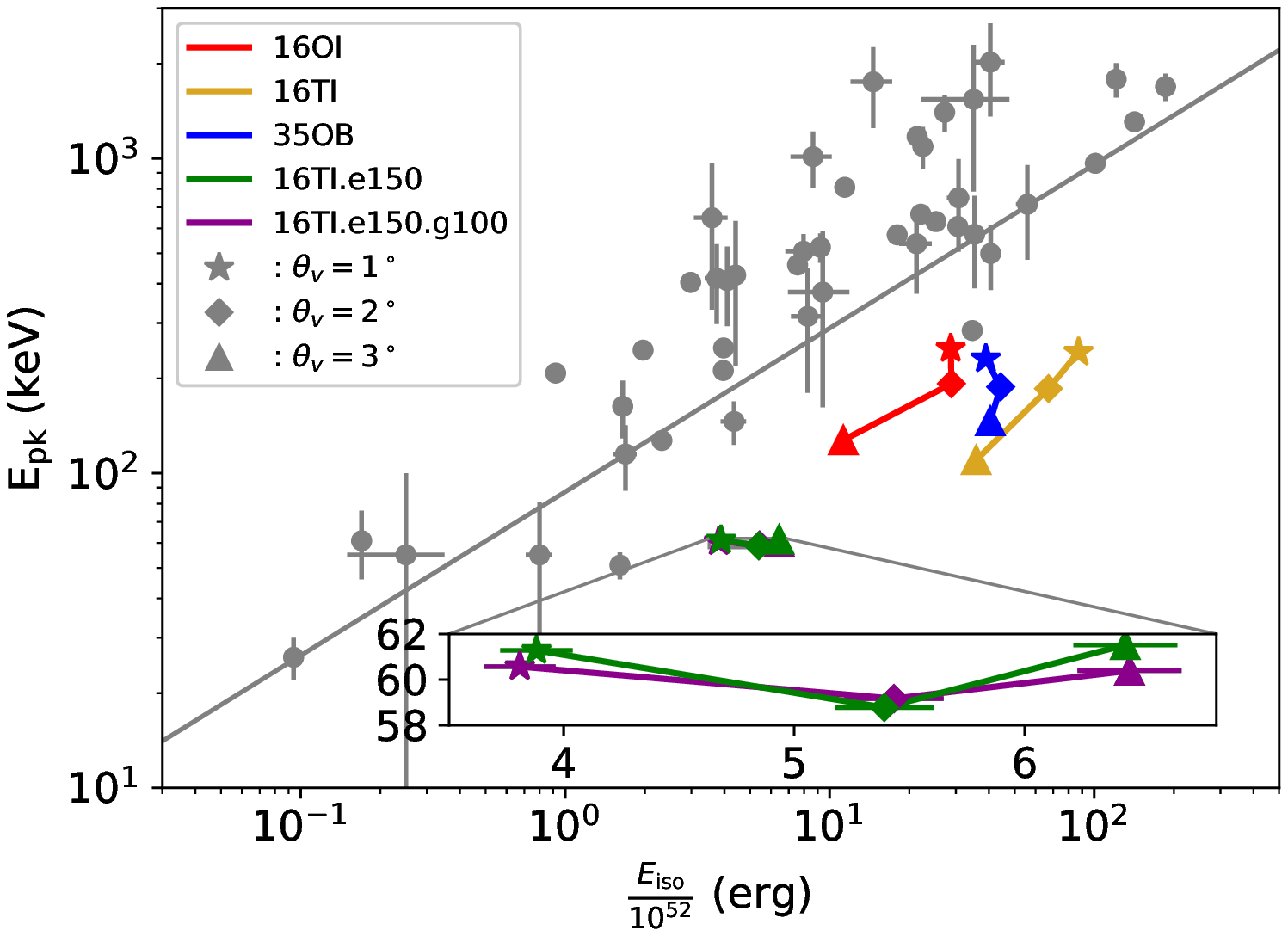}
\caption{The Yonetoku Relation (in the left plot) and the Amati
  relation (in the right plot). The isotropic luminosities and
  isotropic energies are normalized by $10^{52}$. The observational
  data and the fitted relations are plotted in grey as a circle marker
  or a line respectively. The simulation data points are shown in
  color and each viewing angle is differentiated by various marker
  types.  }
\label{y_a_relation}
\end{figure*}

The time-integrated light curve and spectra from our simulations
provide the necessary data for a comparison with GRB ensemble
distributions, such as the Amati and Yonetoku correlations 
\citep{Amati, Yonetoku}. The simulated data is plotted in the
Yonetoku phase space alongside data from \cite{data_set} in
\autoref{y_a_relation}; the gray circles are the observed LGRB data,
the blue line is the fit to the data, the star, diamond, and triangle
markers represent our synthetic data points at
$\theta_\text{v} = 1^\circ, 2^\circ$ and $3^\circ$, respectively. Each
MCRaT simulation is uniquely identified by a different color. It is
clear that all the simulation points lie slightly below the fitted
line. However, the simulation points are well within the spread of the
data and the slope of the correlation is reproduced.

The same simulation results are plotted alongside the Amati
relationship in \autoref{y_a_relation} using the same data set by \cite{data_set}. In this case, the strain with
the observations is more evident, possibly due to en excessive
activity time of the engine in the FLASH simulations (the engine was
active for 100~s, while observations point to a shorter activity time
of $\sim20$ seconds \citep{lazzati_grb_dist,grb_engine_duration}.

\section{Summary and Discussion} 
\label{end} 

We have used the Monte Carlo Radiation Transfer code (MCRaT,
\cite{MCRaT}) to test the photospheric model for a variety of long
Gamma Ray Burst (LGRB) simulations. MCRaT only considers Compton
scattering, as photons are injected and propagated through the LGRB
jet, and does not have the capabilities to investigate non-thermal
particles or synchrotron radiation -- at this time. Even though we
neglect these important mechanisms that affect radiation, we are still
able to reproduce a variety of LGRB characteristics and investigate
their underlying causes.

We show that the
matter and radiation counterparts of the LGRB jet gradually decouple from one another. This allows the photons
to continually interact with the much cooler jet material, thus permitting the average photon energy to decrease even further. 
Photons encounter the photosphere, and are no longer interacting with the jet, when the average temperature of the photons become constant. In this work we also show that the photosphere for a given LGRB moves within a region of space, which
we call the ``photospheric region''. The dynamic nature of the
photosphere is due to the variable density profile in the jet and can affect the spectra of the photons as the photons interact
with the material in the jet for variable periods of time. The work done in this paper exhibits the importance of combining realistic jet dynamics with radiation transport calculations.   

Some populations of photons within our simulation set never reach their respective photospheres
and, as a result, are unable to fully cool to a steady state spectrum. This is due to the small domain of the FLASH simulation
and can be easily remedied with larger domain simulations that allow
the matter to cool for a longer period of time.
  
Additionally, we are able to reproduce the observational values of the
Band $\beta$ parameter in our time resolved simulation spectra. This
gives credence to the photospheric model, although the model still
significantly suffers from a lack of low energy photons, which is
exhibited in our extremely high values of $\alpha$. Our results show
that acquiring Band $\alpha$ parameters that are close to the
observationally expected $\sim -0.75$ is not possible with just
Compton scattering. The peak energies, however, are consistently in
the range of FERMI measurements, with the exception of the simulations
which have lower jet luminosities (i.e. the 16TI.e150 and
16TI.e150.g100 simulations). This success is affected by the fact that
our acquired values of $\alpha$ play a role in the calculation of
$E_{pk}$. Since our values of $\alpha$ are too high, our values for
the $E_{pk}$ become artificially high as well. Thus, the importance of correcting the Band
$\alpha$ parameter increases. In order to correct this, we need to
invoke a photon-producing radiation mechanism that will provide an
extra supply of low energy photons. This is likely where synchrotron
emission will come into play. The region in which these radiation
mechanisms will play a large role is in the sub-photospheric region,
where the radiation and the matter in the jet barely interact,
allowing the low energy photons to escape from the jet after
undergoing a minimal number of interactions without being
thermalized. This would be possible since
\cite{lazzati_variable_photosphere} showed that shocks, which can
reactivate magnetic fields, do extend all the way up to the
photosphere.


By looking at the time-integrated simulation spectra of the 16OI
simulation at $\theta_\text{v} = 1^\circ $, we show that it is
possible to produce a spectrum where the best fit is the Band function
although each time resolved spectrum of the given simulation is best
fit with the COMP function. This effect can be caused by the fact that
many of the time resolved spectra were best fit with the COMP function
due to the exclusion of spectral energy bins in which there were low
photon counts. As the time resolved spectra are added up, the high
energy spectral bins gain enough photons to be included in the fit of
the spectrum, thus permitting the spectrum to be fit by the Band
function in a statistically significant manner. This effect is not
seen in our 35OB simulation at $\theta_\text{v} = 1^\circ $, which
means that some of the spectra may be intrinsically thermal.  The time
resolved spectral fits and light curves for the 16OI, 35OB, and
16TI.e150.g100 simulations are also presented. These light curves
exhibit the photospheric model's ability to recreate some of the
observed relationships between the peak energy and the luminosity. We
can recreate the observed hard-to-soft evolution of $E_{pk}$, and the
tracking behavior between $E_{pk}$ and luminosity. We also see, in
some cases, an anti-correlation between $E_{pk}$ and luminosity which
is not commonly observed.

The simulation set used in this paper are all shown to lie marginally
below the Yonetoku relation, but well within the spread of the data,
which is encouraging. The same simulations are in tension with the
Amati relation. The discrepancy with the observational relations
further show the importance of including a sub-dominant radiation
mechanism in order to correct the spectra. This would correct the Band
$\alpha$ parameter but also push the peak energy of the spectrum
towards lower energies, as \cite{spectral_peak_belo} points out. 
In order to 
correct the low spectral energies, we will need to consider 
sub-photopsheric shocks such as those considered by \cite{Belo_shocks}. These shocks are capable of producing high energy non-thermal particles that will scatter off of photons and increase the average photon energy.
With the production of low energy and high energy photons, it then becomes important to include absorption, which can be facilitated through 
pair production. Emission and absorption processes will be included in future versions of MCRaT, in order to self consistently treat the radiation transfer problem.

As with all hydrodynamic simulations, there is a concern with
resolution being sufficient to resolve small scale features such as
jet re-collimation shocks, which can create non-thermal
particles. This concern is especially serious at large radii, where
the spectrum forms. There is also concern with the number of photons
that we inject into the simulation. The binning of photons in time,
space and energy can quickly whittle down the number of photons
available for spectral analysis thus affecting our spectral fits and
our comparisons with theory. In order to circumvent this problem, we
simply need to inject more photons into the MCRaT simulation, which
will increase the computation time for a given simulation.

\acknowledgements We would like to thank Atul Chhotray, Hirotaka Ito,
and Hiro Nagataki for useful discussions. This work was supported in
part by NASA Swift GI grant NNX15AG96G and ATP grant NNX17AK42G.

\bibliography{references}

\end{document}